\documentstyle [12pt]{article}
\begin{document}
\baselineskip=24.5pt
\setcounter{page}{1}
\thispagestyle{empty}
\topskip 2.5  cm
\vspace{1 cm}

\centerline{\bf Embedding Phenomenological Quark-Lepton Mass Matrices}
\centerline{\bf into SU(5) Gauge Models}
\vskip 1 cm
\centerline{M. Fukugita$^{1,2}$, M. Tanimoto$^3$ and T. Yanagida$^4$}
\vskip 1 cm
\centerline{$^1$ Institute for Cosmic Ray Research, University of Tokyo,
Tanashi, Tokyo 188, Japan}
\centerline{$^2$ Institute for Advanced Study, Princeton, NJ 08540, U. S. A.}
\centerline{$^3$ Faculty of Education, Ehime University, Matsuyama
790-8577, Japan}
\centerline{$^4$ Department of Physics and RESCEU, University of Tokyo,
Tokyo 113-0033, Japan}
\vskip 1 cm
 
\vskip 2 cm
\noindent
{\large\bf Abstract}
\medskip

\noindent
We construct phenomenological quark-lepton mass matrices based on
S$_3$ permutation symmetry in a manner fully 
compatible with SU(5) grand unification.  The Higgs particles we 
need are {\bf 5}, {\bf 45} and their conjugates.
The model gives a charge $-$1/3 quark vs charged lepton mass
relation, and also a good fit
to mass-mixing relations for the quark sector, as well as an attractive
mixing pattern for the lepton sector, explaining a large mixing 
angle between $\nu_\mu$ and
$\nu_\tau$, and either large or small 
$\nu_e-\nu_\mu$ mixing angle, depending on the choice of couplings,
consistent with the currently accepted 
solutions to the solar neutrino problem.

\newpage
\topskip 0.0 cm
\noindent

Predictions of the quark-lepton mass spectrum are the least successful
aspect of the unified gauge theories. In classical SU(5) grand 
unification (GUT) with the simplest choice of Higgs scalars, we obtain the mass
relations between the charge -1/3 quarks (referred simply to as 
down quarks) and charged leptons, 
$m_d=m_e$, $m_s=m_\mu$ and $m_b=m_\tau$. While the last of these
relations agrees with experiment [1], the other two are far from
the reality. Georgi and Jarlskog (GJ) [2] have shown that 
the mass degeneracy of
$d/e$ and $s/\mu$ can be lifted by introducing {\bf 45}-plet Higgs
to give at the GUT scale
$$m_d=3m_e,~~~~~ m_s=(1/3)m_\mu,~~~~~ m_b=m_\tau \eqno{(1)}$$
in reasonable agreement with experiment.
No prediction, however, has been given within the SU(5) framework to
the charge 2/3 quark spectrum and hence to quark mixing  which
is usually described in terms of the 
Cabibbo-Kobayashi-Maskawa (CKM) matrix. By extending
the unifying group to SO(10), one may relate the charge 2/3 quark 
(referred to as up quark) mass to
the Dirac mass of neutrinos, and also the quark mixing angles to the lepton
mixing angles. The prediction, however, is too tight, and does not
seem to be consistent with experiment (e.g. [3]),
or else the Higgs content becomes very complicated [4].

There is an empirical approach successful
in giving the quark mass-mixing relation at a phenomenological level.
It usually assumes some {\it ad hoc} symmetry imposed on the mass matrices,
as advocated first by Fritzsch [5]. Several simple 
representations of the quark mass matrices have been known that
give mass-mixing relations in fair agreement with experiment [6]. 
This approach also successfully applies to understanding the 
relation between lepton mixing and neutrino mass [7] indicated by
the solar neutrino problem [8] and by the atmospheric neutrino 
experiment [9]. In ref.[10] we have proposed specific quark-lepton 
mass matrices based on 
S(3)$_L\times$S(3)$_R$ symmetry, or so-called ``democratic'' principle [11], 
and small symmetry breaking terms, which lead to  
attractive mixing patterns for neutrinos.

One of the problems with the ``phenomenological matrix approach'' is that
the consistency with the unified gauge model is unclear. If one
straightforwardly imposes the compatibility with a gauge model, on the other
hand, we are 
usually led to unwanted relations for quark and lepton masses as remnants of 
prototype gauge models.

In this paper we show that there exists a successful matrix model based on
the S$_3$ symmetry approach, which is fully
compatible with SU(5) GUT's and at the same time gives
predictions for quark and lepton masses and their mixings that agree 
with experiment [12]. This compatibility 
reduces the arbitrariness of the matrices, but also inspires us to modify
them for the quark sector [13], 
which brings the predicted quark mixing
in good agreement with experiment,
especially for the (2,3) sector of flavour.
For the neutrino sector all results presented in ref.[10] 
are retained.

We start with the observation that the phenomenological matrices given
in [13] (for quarks) and [10] (for leptons) are compatible with the
Yukawa couplings in the 
presence of a {\bf 5}-plet Higgs of SU(5); an introduction of a {\bf 45}-plet,
which is necessary to lift unwanted mass degeneracy, requires only a minimum
modification of the symmetry breaking matrix for the up quark sector.
We write the Higgs coupling  
$${\cal L}_{\rm Yukawa}= Y(5_H)_{Uij} {\bf 10}_i {\bf 10}_j  {\bf 5}_H +
   Y(45_H)_{Uij} {\bf 10}_i {\bf 10}_j  {\bf 45}_H + $$
$$  + Y(5_H^*)_{{D/E}ij} {\bf 5}^*_i {\bf 10}_j {\bf 5}_H^* +
   Y(45_H^*)_{{D/E}ij} {\bf 5}^*_i {\bf 10}_j {\bf 45}_H^* + 
     \kappa(5_H 5_H)_{\nu ij} {\bf 5}^*_i {\bf 5}^*_j 
   {{\bf 5}_H{\bf 5}_H \over M_R}
                                      \eqno{(2)}$$
where bald face symbols 
with suffix $H$ denote Higgs scalars of a specified multiplet,
and those with suffix $i$ or $j$ (refer to flavour) 
are SU(5) matter fields,
${\bf 5}_i =(d^c_1, d^c_2,d^c_3, e^-, \nu_e)_{Li}$ and
${\bf 10}_j =(u^c_1, ..., u^c_1, ..., d^c_i, ..., e^+)_{Lj}$.
We write the down quark and charged lepton sectors as $D/E$ as they 
are unified.
The last term of eq. (2) is an effective neutrino coupling where 
the neutrino is taken to be of the Majorana type. We suppose that the main part
of the mass term arises from {\bf 5}$_H$;  {\bf 45}$_H$ gives only  
perturbations.

We postulate that the main part of the mass matrices is 
S$_3$ permutation symmetry invariant,
i.e., S$_3^{\bf 10}\times$S$_3^{\bf 5}$ in our context ({\bf 10} and {\bf 5}
refer to representations of fermions).
The choice of the Yukawa coupling matrix 
$Y(5_H^*)_{D/Eij}$  is then unique: 
$$Y =      \left[ \matrix{1 & 1 & 1 \cr
                            1 & 1 & 1 \cr
                            1 & 1 & 1 \cr
                                         } \right]  . \eqno{(3)}$$
Two matrices are allowed for $Y(5_H)_{Uij}$ from the invariance under 
S$_3^{\bf 10}\times$S$_3^{\bf 5}$ ,
$$      \left[ \matrix{1 & 0 & 0 \cr
                            0 & 1 & 0 \cr
                            0 & 0 & 1 \cr
                                         } \right]  ,~~~~~~~~~~~~~~~~~  
             \left[ \matrix{0 & 1 & 1 \cr
                            1 & 0 & 1 \cr
                            1 & 1 & 0 \cr
                                         } \right].  \eqno{(4)}$$
We take the 1:1 combination of the two to give form
(3) for simplicity and for agreement with the democracy argument as in
[11, 13]. We also assume for $ \kappa(5_H5_H)_{ij}$ 
the first matrix of (4) for the reason explained in
[10].

We break S$_3^{\bf 10}\times$S$_3^{\bf 5}$ symmetry with an extra Yukawa
coupling to
{\bf 5}$^*_H$ and a coupling to 
{\bf 45}$^*_H$ for 
down-quark and charged-lepton sectors, and  
{\bf 5}$_H$ and {\bf 45}$_H$ for the up-quark sector. 
Namely, our mass matrices
are
$$M_D=[Y(5_H^*)_D + a Y(45_H^*)_D] \langle \phi_{5_H}^* \rangle , \eqno{(5)}$$
$$M_E=[Y(5_H^*)_D - 3 a Y(45_H^*)_D] \langle \phi_{5_H}^*\rangle  , \eqno{(6)}$$
$$M_U=[Y(5_H)_U + b Y(45_H)_U] \langle \phi_{5_H} \rangle , \eqno{(7)}$$
where $a= \langle \phi_{45_H}^* \rangle_D/\langle \phi_{5_H}^* \rangle_D$ and
$b= \langle\phi_{45_H} \rangle_U/\langle \phi_{5_H} \rangle_U$.  Or,
more explicitly, we write
$$M_D= {K \over 3}\Bigg(
\left[ \matrix{1 & 1 & 1 \cr
                            1 & 1 & 1 \cr
                            1 & 1 & 1 \cr
                                         } \right]
+\left[ \matrix{-\epsilon_D & 0 & 0 \cr
                         0 & \epsilon_D & 0 \cr
                         0 & 0 & \delta_D \cr
                                              } \right] \Bigg)  \eqno{(5')}$$
for the down-quark sector. 
At this level the symmetry breaking term of (5$'$) may
be ${\bf 5}^*$ or ${\bf 45}^*$.
The Higgs leading to symmetry breaking 
is either new Higgs that develops a small vacuum
expectation value or the same Higgs giving the main mass term but
with small Yukawa couplings. We  do not
distinguish these two possibilities here.

For the charged lepton sector,
 $$M_E= {K \over 3}\Bigg(
\left[ \matrix{1 & 1 & 1 \cr
                            1 & 1 & 1 \cr
                            1 & 1 & 1 \cr
                                         } \right]
+\left[ \matrix{-\epsilon_L & 0 & 0 \cr
                         0 & \epsilon_L & 0 \cr
                         0 & 0 & \delta_L \cr
                                              } \right] \Bigg) . \eqno{(6')}$$
If all symmetry breaking terms in the second matrices of ($5'$) and ($6'$)
come from ${\bf 5^*_H}$ (i.e., $a=0$) 
we are led to unwanted down quark-charged lepton 
mass degeneracy.
This problem can 
be avoided by assuming that $\delta$ elements are generated 
from the coupling to a
${\bf 45^*}_H$-plet Higgs scalar, while $\epsilon$ terms are 
from ${\bf 5^*}_H$. We have then
$$\epsilon_D=\epsilon_L=\epsilon     \eqno{(8)}$$
and 
$$\delta_D=-\delta_L/3 =\delta,      \eqno{(9)}$$
with real $\epsilon$ and $\delta$ [14].
We obtain
$$ m_b\simeq K(1+\delta/9), ~~~
  m_s\simeq 2K \delta /9, ~~~
  m_d\simeq -K \epsilon^2(1+\delta)/6\delta ,       \eqno{(10)}$$
$$  m_\tau\simeq K(1-3\delta/9),~~~
    m_\mu\simeq -2K\delta/3, ~~~
    m_e\simeq K\epsilon^2(1-3\delta)/18\delta ,    \eqno{(11)}$$
after diagonalization of the matrices. The masses of (10) and (11)
satisfy the SU(5) GJ mass relation (1) when $\delta\ll 1$. 
Now all parameters in the down-quark and charged-lepton sectors are 
determined solely by $e,\mu$ and $\tau$ masses. 

If we would take the same form as (5$'$) also
for the up-quark sector [13], i.e., the 
symmetry breaking term necessarily limited to ${\bf 5^*}$,  
we were led to $V_{23}\simeq
0.015$ compared with experiment $0.036- 0.042$, whereas $V_{12}$ is 
successfully predicted. We note, however, in our scheme (5-7) that
the breaking terms for up- and down-quark sectors may not necessarily be
the same form. In fact,  ${\bf 45}_H$ Higgs, when coupled to
${\bf 10}_i\times {\bf 10}_j$, should give matrix elements  
different from (5$'$). That is, it gives rise to
flavour off-diagonal elements,
rather than diagonal due to the antisymmetric nature of ${\bf 45}_H$. 
Therefore,
we  take for (7),
$$M_U= {K' \over 3}\Bigg(
\left[ \matrix{1 & 1 & 1 \cr
                            1 & 1 & 1 \cr
                            1 & 1 & 1 \cr
                                         } \right]
+\left[ \matrix{-\epsilon_U & 0 & \delta_U \cr
                         0 & \epsilon_U & \delta_U \cr
                         -\delta_U & -\delta_U & 0 \cr
                                              } \right] \Bigg)  \eqno{(12)}$$
where $\delta_U$  comes from {\bf 45}$_H$. $\delta_U$ in the (1,3)
matrix element may generally differ from that in (2,3), or simply it 
may even vanish. We take a parallelism with the down quark sector
that {\bf 45}$^*_H$ couples to the third generation matrix elements:
we found that the simple choice taken here
gives resulting mixing angles in agreement with experiment.

For this mass matrix the mass eigenvalues are ($\delta_U$ and $\epsilon_U$ 
being real)
$$m_t\simeq  K'(1-2\delta_U^2/9),~~~     
  m_c\simeq  2K'\delta_U^2/9,~~~
  m_u\simeq  -K'\epsilon_U^2/6\delta_U^2,            \eqno{(13)}$$
and resulting quark mixing angles read 
$$V_{12}\simeq \sqrt{m_s/m_d}-\sqrt{m_u/m_c}   \eqno{(14)}$$
and
$$V_{23}\simeq m_s/\sqrt2 m_b -\sqrt{m_c/m_t}  \ . \eqno{(15})$$
The explicit analytical expression is more complicated for $V_{13}$,
and we do not bother the reader by writing it here.

Let us now carry out a numerical analysis.
The parameters for the down-quark and charged-lepton sectors are fixed by
$m_e, m_\mu, m_\tau$ to be $K=1.13$ GeV, $\epsilon=0.019$ and
$\delta=-0.093$. Here, we should use masses at GUT energy scale,
and our input parameters are taken from two-loop calculations of 
ref. [15] as presented in Table. 1, where
we also compare predicted down-quark masses
with those expected at GUT mass scale.
The agreement of the
prediction with ``experiment'' is good, though $m_d$ is 
somewhat smaller [16].
For the up-quark
sector we take $K'=129$ GeV, $\epsilon_U=-0.00072$ and $\delta_U=-0.103$
to fit the central values of $u,c,t$ masses in Table 1.

The resulting mixing matrix is
$$|V_{\rm quark}|= 
             \left[ \matrix{0.975 & 0.220 & 0.0086 \cr
                           0.220 & 0.975 & 0.036 \cr
                           0.016 & 0.033 & 0.999 \cr
                                         } \right]  \eqno{(16)}$$
which is compared with experimental values $|V_{12}|=0.217-0.224$,  
$|V_{23}|=0.036-0.042$
and $|V_{13}|=0.002- 0.005$.  
We emphasize that an excellent agreement is achieved with experiment
for $V_{23}$, whereas a factor of two disagreement has been 
taken to be a problem with the democratic matrix [13].
Our solution seems to be the simplest among others [6].
While the predicted $V_{13}$ is somewhat larger
than experiment, we do not pursue this problem further here [we can
bring $V_{13}$ in a good agreement with experiment 
without conflicting with our
principles, if $\epsilon_D$ is put on the (1,3) and (3,1) components,
in addition to (1,1) and (2,2), of
(5$'$), see [17]].

The Majorana neutrino coupling to Higgs is taken freely 
from the other  
sectors, since therein ${\bf 5}^*_i {\bf 5}^*_j$ does not appear.
The only
requirement is that the matrix should respect 
S$_3^{\bf 10}\times$S$_3^{\bf 5}$ in its main part, and we take 
the first matrix of (4). We may take the S$_3^{\bf 10}\times$S$_3^{\bf 5}$
breaking terms to be
$$M_\nu^{(1)}=\left[ \matrix{0 & \epsilon_\nu & 0 \cr
                 \epsilon_\nu  & 0 & 0 \cr
                  0 & 0 & \delta_\nu \cr} \right],\eqno{(17)}$$
or
$$M_\nu^{(1)}=\left[ \matrix{-\epsilon_\nu & 0 & 0 \cr
                 0 & \epsilon_\nu  & 0  \cr
                  0 & 0 & \delta_\nu \cr} \right].\eqno{(18)}$$
This is exactly the same as the model we have presented in [10],
and the lepton (neutrino) mixing matrix reads 
$$|V_\ell|=\left[ \matrix{0.998 & 0.045 & 0.050 \cr
                        0.066 & 0.613 & 0.787 \cr
                        0.005 & 0.789 & 0.614 \cr
                                         } \right], \eqno{(19)}$$
or
$$|V_\ell|\simeq \left[ \matrix{0.737 & 0.674 & 0.050 \cr
                  0.386 & 0.479 &  0.787 \cr
                           0.555 &  0.562 &  0.614 \cr
                                         } \right], \eqno{(20)}$$
according as we take (17) or (18). The second mixing matrix 
virtually agrees
with the one given by Fritzsch and Xing [7] derived from 
different principles.  
The mass eigenvalues are
 $K_\nu\pm\epsilon_\nu$, and 
$K_\nu+\delta_\nu$, hence describing three neutrinos almost degenerate
in mass. We note that the mixing matrices are predominantly determined by
charged lepton masses; neutrino masses change the elements little.
The case with (17) describes
large mixing for $\nu_\mu-\nu_\tau$, 
$\sin^2 2\theta_{\mu\tau}\simeq 0.93 $ or the $\nu_\mu$ survival fraction of 
54\% in the atmospheric neutrino experiment, and small mixing for
$\nu_e-\nu_\mu$ ($\sin^2 2\theta_{e\mu}\simeq 8\times 10^{-3}$) consistent
with the small angle solution of the MSW explanation for the solar 
neutrino problem [8]. The case with (18) predicts large mixing for both
$\nu_\mu-\nu_\tau$ and $\nu_e-\nu_\mu$, the latter being consistent
with either MSW large angle solution or mixing angle required in
solar neutrino oscillation in vacuum, independent of the neutrino 
mass difference squared. The (1,3) elements of (19) and (20) come out
to be consistent with the CHOOZ experiment [18], which yields 
roughly $<0.2-0.3$
for this element when the mass of $\nu_\tau$ is in the range 
Super-Kamiokande indicates.

In conclusion, we have shown that one could successfully embed 
phenomenological quark-lepton mass matrices obtained by the democracy
principle into the SU(5) scheme. The choice of matrices is not yet
unique, but the
interlocking of the two principles
tightly constrains the allowed form, and reduces the
number of parameters of the model; the matrix for charged leptons
is no longer independent of that for down quarks. In addition we  
have found 
the matrices in better agreement with experiment after tweaking
to reconcile with the SU(5)
than other empirical
matrices constructed without referring to gauge models. 
The approach
reconciling empirical matrices with gauge theory as we have done in this paper
might perhaps give a guiding principle to understand the Higgs sector of
gauge theories, which otherwise appears too arbitrary.

{\bf Note added}: we have learned that Mohapatra and Nussinov [19] 
have recently
proposed a gauge model for S$_3$ symmetric mass matrices embedded into
SU(2)$_L\times$SU(2)$_R\times$U(1)$_{B-L}$.

\vskip10 mm
\noindent
{\bf Acknowledgements}

The authors are supported by Grants-in-Aid of the Ministry of Education.

\vskip 2 cm

\newpage

\noindent
Table 1. Input quark-lepton mass parameters and the prediction of our
model (mass in MeV units). ``exp'' means the experimental vale expected 
at GUT energy scale, 
as given by a two loop analysis of [15]. The input
masses given here are used to predict the CKM matrix (16) and the lepton
mixing matrix (19) or (20).

\tabskip=1.5mm plus 1mm minus 1mm
\halign to \hsize{
\hfil # \hfil & \hfil # \hfil & \hfil # \hfil & \hfil # \hfil &
\hfil # \hfil & \hfil # \hfil & \hfil # \hfil & \hfil # \hfil &
\hfil # \hfil & \hfil # \hfil \cr
\noalign{\vskip 3mm \hrule\vskip 0.1mm}
\noalign{\vskip 1mm \hrule\vskip 3mm}
     &   $m_e$ &   $m_\mu$ &   $m_\tau$ &   
$m_d$ &     $m_s$ &   $m_b$  &  $m_u$  & 
  $m_c$ &  $m_t$ \cr
\noalign{\vskip 3mm \hrule\vskip 3mm}
``exp.'' & $0.325$ & $68.60$  & $1171$ & $1.3\pm 0.2$ &  $26.5 ^{+3.4}_{-3.7}$  
& $1000\pm 40$ & 
$1.0\pm 0.2$ & $302^{+25}_{-27}$ & $129^{+196}_{-40}\times10^3$ \cr
pred. & input  & input & input & 0.67 & 24.4 & 1120 & 
input & input & input \cr
\noalign{\vskip 3mm \hrule\vskip 3mm}
}

\end{document}